\documentclass[superscriptaddress,preprint,nofootinbib,nobibnotes,eprint,amsmath,amssymb,aps,prb,citeautoscript,titlepage,nobalancelastpage,floats,final,eqsecnum]{revtex4-2}

\usepackage[utf8]{inputenc}

\usepackage{amsmath}
\usepackage{amssymb}
\usepackage{amsfonts}

\usepackage{latexsym}
\usepackage{slashed}
\usepackage[svgnames]{xcolor}
\usepackage{soul}
\usepackage{braket}
\usepackage{wasysym}
\usepackage{mathrsfs}  
\usepackage{physics}
\usepackage{enumitem}        

\usepackage{graphics}
\graphicspath{ {graphs/} }
\usepackage{booktabs}

\usepackage{microtype}
\usepackage{multirow}

\usepackage{hyperref}

\usepackage{mathtools}

\usepackage{easyReview}

\begin{document}
	
	\title{\Large{{\sc  Non-Abelian Zero Modes in Fractional Quantum Hall-Superconductor Heterostructure}}}

	\author{Gustavo M. Yoshitome}
	\email{gustavo.yoshitome@uel.br }
	\affiliation{Departamento de Física, Universidade Estadual de Londrina, Londrina, PR, Brasil}

	\author{Pedro R. S. Gomes}
	\email{pedrogomes@uel.br}
	\affiliation{Departamento de Física, Universidade Estadual de Londrina, Londrina, PR, Brasil}

	
\begin{abstract}	
	
We discuss the emergence of non-Abelian zero modes from twist defects in Abelian topological phases. We consider a setup built from a fractional quantum Hall (or a fractional Chern insulator)-superconductor heterostructure, which effectively induces a phase transition, leading to a topological phase endowed with new anyonic symmetries, and accordingly supporting distinct types of zero modes at fixed filling. These defects are modeled at the interface between two copies of the same heterostructure arranged side by side, which produces counterpropagating modes that can be gapped by interactions that realize the anyonic symmetries. We characterize the parafermions associated with each anyonic symmetry and discuss how their presence affects the periodicity of the Josephson tunneling current.
	
\end{abstract}

	\maketitle

\tableofcontents
	

\section{Introduction}

One of the greatest challenges in the field of topological phases of matter is the detection of the elusive non-Abelian emergent excitations. Despite intense efforts, the experimental front is facing tremendous difficulties. The provision of controllable physical setups that are able to support such exotic excitations is thus of great relevance for the field. In this spirit, non-Abelian excitations have been predicted to manifest in Abelian phases in the form of defects associated with anyonic symmetries, which correspond to permutations of anyons preserving the fusion algebra and the braiding statistics \cite{Barkeshli:2014cna,Khan:2014waa,Teo:2015xla,Teo_2016}.

A well-established system exhibiting an anyonic symmetry is the Laughlin phase of the fractional quantum Hall (FQH) effect with filling $\nu=1/k$, where $k$ is an odd integer. For $k>1$, this phase has $k$ nontrivial anyons, which are permuted by the action of charge conjugation. Specifying the anyons in terms of integers $n~ \text{mod} ~k$, the statistics of the anyon $n$ is $\nu=n^2/k~\text{mod}~2$. Charge conjugation acts on the anyons as $n \rightarrow k-n$, under which the statistics changes as $n^2/k\rightarrow n^2/k+k$. Since the anyons $n$ and $k-n$ differ only by the transparent excitation (fermion), this permutation preserves the fractional quantum numbers, corresponding to an anyonic symmetry.

In an insightful work \cite{Clarke_2013}, the authors implicitly explore the anyonic symmetry of the Laughlin phase to conceive a concrete physical setup where the non-Abelian zero modes manifest in an explicit way, generating potentially detectable physical signatures. The essential idea is to use the boundary to realize the presence of defects. To this end, they consider two Laughlin phases arranged side by side. At the one-dimensional interface between the phases, superconducting and insulating potentials are induced in alternating regions, producing domain walls. The crucial point is that the superconductor effectively implements charge conjugation, as an electron of charge 1 crossing the interface through the superconducting region has its charge modified to $-1$ mod 2 (i.e., modulo the Cooper pair charge). The domain walls separating the superconducting and insulating regions correspond to the locations of the defects, which host non-Abelian zero modes. Subsequently, several works considered the emergence of parafermions in similar setups \cite{Lindner_2012,Cano_2015,Ebisu_2017,Santos_2017,May_Mann_2019,bollmann2025phasesquasionedimensionalfractionalquantum,Burrello_2013,Chen:2015blr,Mong_2014}.

In this work, we pursue this program of exploring anyonic symmetries in Abelian phases to produce non-Abelian zero modes. Specifically, we consider a heterostructure composed of a Laughlin phase (or a Chern insulator) at filling $\nu=1/k$ in proximity to a conventional $s$-wave superconductor in the strong-coupling regime, as shown in Fig. \ref{Heterostructure}\footnote{We emphasize that the whole Laughlin phase (bulk+edge) is in proximity to the superconducting potential. This is different from the setup discussed in \cite{Clarke_2013}, where only the boundary is in proximity to the superconductor.}. This type of setup has been discussed previously in Refs. \cite{Qi_2010,Vaezi_2013,Khan_2017,Gomes:2022lwl}\footnote{The recent work \cite{kudo2025} considers a similar setup involving the proximity of a composite Fermi liquid and a conventional superconductor.}. In particular, as pointed out in \cite{Khan_2017}, the presence of the superconductor introduces additional fractionalization of the quantum Hall anyons, in such a way that the fermion is no longer the transparent excitation. The local excitation turns out to be the Cooper pair. In effect, this reconfiguration of the spectrum of anyons of the composite system ends up introducing several anyonic symmetries, which gives the possibility of realizing different types of non-Abelian zero modes at fixed filling.

One delicate point in using boundaries to realize defects associated with anyonic symmetries is that the boundary itself contributes nontrivially to the ground-state degeneracy, mingling with the contributions from the anyonic symmetries. A simple case illustrating this point is obtained by setting $k=1$ (the integer quantum Hall (IQH) limit) in the discussion of \cite{Clarke_2013}. In this case, they predict the presence of Majorana zero modes trapped in the domain walls. As the IQH state does not have any nontrivial anyonic symmetries, we see that these zero modes are intrinsic to the boundary. This intrinsic boundary degeneracy corresponds to the usual degeneracy in one-dimensional fermionic systems \cite{Kitaev_2001,PhysRevB.83.075103,PhysRevB.84.195436}. Nevertheless, we shall discuss how to disentangle the contribution of the anyonic symmetry from the contribution of the intrinsic boundary physics.

Previous works have developed explicit realizations of non-Abelian zero modes in a variety of related systems, including Refs.~\cite{Clarke_2013,Lindner_2012,Cano_2015,Ebisu_2017,Santos_2017,May_Mann_2019,bollmann2025phasesquasionedimensionalfractionalquantum,Burrello_2013,Chen:2015blr,Mong_2014}. In parallel, anyonic symmetries and their associated twist defects have been studied from a complementary perspective, as in Ref.~\cite{Khan_2017}. The main novelty of our work is to unify these two lines of development: we provide a systematic framework that starts from bulk anyonic symmetries and leads to an explicit boundary realization of the associated non-Abelian zero modes. In doing so, we carefully account for the subtleties involved in separating bulk and boundary contributions to the defects. We further derive physical signatures of these zero modes in Josephson currents.

To be concrete, we focus on the case $k=3$, but there is no conceptual difficulty in generalizing to the case of arbitrary $k$. The case $k=3$ exhibits three anyonic symmetries, referred to as charge conjugation, fermion parity flip, and composite symmetry (the composition of the previous ones). We then proceed with the study of the physical setup able to support the defects, which trap $\mathbb{Z}_6$ non-Abelian zero modes in the case of charge conjugation, $\mathbb{Z}_2$ zero modes in the case of fermion parity flip, and $\mathbb{Z}_3$ zero modes in the case of the composite symmetry. In addition, we construct effective Hamiltonians that dispose of the intrinsic boundary contributions and use them to derive physical signatures of the zero modes in the periodicity of the Josephson tunneling current.

This work is organized as follows. In Section \ref{II}, we show the effective topological field theory for the superconductor-FQH heterostructure and its anyonic content. In Sec. \ref{III}, we show the anyonic symmetries resulting from this topological order. We discuss how to understand the corresponding symmetries from the point of view of boundary phases by folding the bulk in Sec. \ref{IV}. In Sec. \ref{V}, we show how to construct the parafermion zero-mode operators that emerge at the domain walls. In Sec. \ref{VI}, we derive the signatures of these parafermions in the periodicity of the Josephson tunneling current.


\section{Quantum Hall-Superconductor Heterostructure \label{II}}

\begin{figure}
	\includegraphics[scale=.5,angle=90]{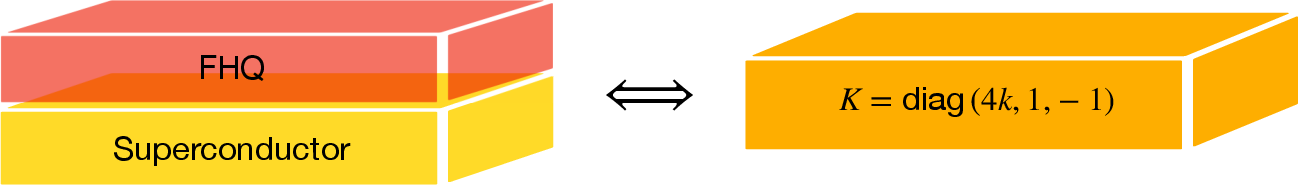}
	\caption{Quantum Hall-superconductor heterostructure, which is equivalent to a topological phase described by the $K$-matrix $K=\text{diag}(4k,1,-1)$.} 
	\label{Heterostructure}
\end{figure}

We consider a layered heterostructure formed by a fractional quantum Hall (FQH) phase and a conventional $s$-wave superconductor, as shown in Fig.~\ref{Heterostructure}. Following Ref.~\cite{Hansson_2004}, we treat the superconductor as a topologically ordered phase, in which half-flux vortices are deconfined\footnote{As explained in Ref.~\cite{Khan_2017}, the system with deconfined vortices has the same anyonic symmetry content as the case in which the vortices are confined.}. In this situation, the low-energy topological response of the coupled system is captured by the effective action
\begin{equation}
S=\int \frac{k}{4\pi} \alpha d\alpha  -  \frac{1}{2\pi } a d\alpha + \frac{2}{2\pi} adb ,
\label{composite}
\end{equation}
where $\alpha$ is the emergent gauge field of the Laughlin FQH phase, while $a$ and $b$ are the gauge fields of the superconducting BF theory. The term $(k/4\pi)\alpha d\alpha$ describes the intrinsic topological order of the FQH state, with the odd integer level $k$ determining the filling fraction. The BF term $(2/2\pi)adb$ encodes the mutual braiding between superconducting charges and half-flux vortices. Finally, the mixed term $-(1/2\pi)a d\alpha$ represents the electromagnetic coupling between the FQH charge sector and the superconducting charge sector. Thus, the strong coupling between the FQH phase and the superconductor is implemented by promoting the electromagnetic probe field of the FQH response to the dynamical gauge field $a$ of the superconducting topological field theory.


The action \eqref{composite} can be written in a $K$-matrix form
\begin{equation}
	S= \int \frac{K_{IJ} }{4\pi} a_I d a_J, ~~~I,J=1,2,3,
	\label{kmatrix}
\end{equation}
where $(a_1,a_2,a_3)\equiv (\alpha, a, b)$ and 
\begin{equation}
	K=\begin{pmatrix}
		k&-1& 0
		\\
		-1&0&  2
		\\
		0&2& 0 \\
	\end{pmatrix}.
\end{equation}
The topological phase described by this $K$-matrix can be transformed into an equivalent one through $a\rightarrow W^T a$, where
\begin{equation}
	W=\begin{pmatrix}
		2&2k& 1
		\\
		1&\frac{k-1}{2}&  0
		\\
		-1&-\frac{k-1}{2}& 0 \\
	\end{pmatrix},~~~ \text{det}(W)=-1,
\end{equation}
with $W\,\in\, GL(3,\mathbb{Z})$ (recall that $k$ is odd). Under the transformation $K\rightarrow W K W^{T}$, we obtain the new $K$-matrix
\begin{equation}
	K=\begin{pmatrix}
		4k&0& 0
		\\
		0&1&  0
		\\
		0&0& -1 \\
	\end{pmatrix}.
	\label{diagonalizing}
\end{equation}
With the diagonal form in Eq.~\eqref{diagonalizing}, the physical properties of the topological phase become particularly transparent: the theory separates into a $U(1)_{4k}$ sector and a $(1,-1)$ local fermionic sector. The nontrivial anyonic content relevant for the twist defects is encoded in the $U(1)_{4k}$ sector, while the  $(1,-1)$ sector accounts for the local fermionic degrees of freedom. We emphasize, however, that this separation should be understood as a convenient basis choice for the present theory. In a generic Abelian $K$-matrix theory, such a diagonalization may not be possible under allowed integral basis transformations $K\to WKW^T$, with $W\in GL(N,\mathbb Z)$, and we may not be able to isolate a single topological sector in this way. Thus, while the decomposition into $U(1)_{4k}$ and $(1,-1)$ sectors is basis-dependent, the anyonic symmetry data and the associated defect structure are basis-independent.


The emergent quasiparticles are described by line operators
\begin{equation}
\mathcal{W}_{\bf l}	\equiv e^{i\oint l_I a_I},~~~ {\bf l} \in \mathbb{Z}^3,
\label{wl}
\end{equation}
where $\mathbb{Z}^3$ stands for a 3-vector with integer components, as required by invariance under large gauge transformations. Two quasiparticles are equivalent if they differ by a local quasiparticle, i.e., ${\bf l} \sim {\bf l} + K \,\mathbb{Z}^3$.  The self-statistics and mutual statistics of the quasiparticles described by the line operators in \eqref{wl} are given by
\begin{equation}
\delta_{\bf l}=   \pi {\bf l}^T K^{-1} {\bf l}~~~\text{and}~~~ \theta_{{\bf l},{\bf m}}= 2 \pi {\bf l}^T K^{-1} {\bf m},
\label{statistics}
\end{equation}
respectively. 


\section{Anyonic Symmetries and Defects \label{III}}

Anyonic symmetries correspond to permutations of anyons preserving the fusion algebra and the braiding statistics \cite{Barkeshli:2014cna,Teo:2015xla}. For the Abelian case, this translates into the permutations of anyons that preserve the relations in \eqref{statistics}. In the $K$-matrix formulation, the anyonic symmetries act on the fields as $a\rightarrow W^T a$, under which $K\rightarrow W K W^T$, with the condition
\begin{equation}
	W K W^T=K,~~~ \abs{\det W}=1.
	\label{as}
\end{equation} 

To unveil the symmetry matrices $W$, it is helpful to notice that the fermionic sectors of the phase described by \eqref{diagonalizing} do not carry nontrivial anyonic symmetries. They reside exclusively in the bosonic sector. Thus we shall look for transformations that keep invariant   
\begin{equation}
\delta_{l_1,0,0}=  \frac{l_1^2}{4k}~\,\text{mod}~2.
\end{equation}
From this expression, we see that there are two anyonic symmetries that are quite general in the sense that they are present for every level $k>1$. The first one is the {\it charge conjugation}, which acts simply as $l_1 \rightarrow -l_1$. The second symmetry maps all lines with odd $l_1$ into $l_1\rightarrow l_1 +2k$. Under this shift,
\begin{equation}
	\frac{l_1^2}{4k}~~\rightarrow~~ \frac{l_1^2}{4k} + (l_1+k)= \frac{l_1^2}{4k} ~\,\text{mod}~2,
\end{equation}
where the last equality follows only for $l_1$ odd. This symmetry acts trivially on the lines with even $l_1$. It is denoted as the {\it fermion parity flip} \cite{Khan_2017}. The composition of charge conjugation and fermion parity flip gives rise to another anyonic symmetry. 

In general, a mapping $l_1 \rightarrow p\,l_1$ corresponds to a symmetry for an integer $p$ satisfying
\begin{equation}
	p^2 =1~\,\text{mod}~8k. 
\end{equation}
We note that for higher values of the level $k$, it may happen that the phase possesses additional level-dependent symmetries. For example, for $k=15$, there is an additional symmetry with $p=11$, which does not correspond to any of the symmetries discussed above.

Now, we need to investigate the implementation of these symmetries in the $K$-matrix theory \eqref{diagonalizing}, i.e., we need to find matrices $W$ leading to the above transformations for $l_1$ and satisfying the constraints in \eqref{as}. 

\subsubsection{Case $k=3$}

To proceed, we shall consider the concrete case of $k=3$, in which $K=\text{diag}(12,1,-1)$. For this case, there are no other symmetries besides the ones discussed previously. A representative set of $W$-matrices is given as follows. Charge conjugation can be implemented by the matrix 
\begin{equation}
	W_{\text{cc}}=\begin{pmatrix}
		-1&0& 0
		\\
		0&-1&  0
		\\
		0&0& -1 \\
	\end{pmatrix},~~~ \text{det}(W_\text{cc})=-1,~~~ W^{-1}_{\text{cc}}=W_{\text{cc}}.\label{cc}
\end{equation}
Fermion parity flip  can be implemented by 
\begin{equation}
	W_{\text{fpf}}=\begin{pmatrix}
		7&0& -24
		\\
		0&-1&  0
		\\
		2&0& -7 \\
	\end{pmatrix},~~~ \text{det}(W_{\text{fpf}})=1,~~~ W_{\text{fpf}}^{-1}=W_{\text{fpf}}.
	\label{parity}
\end{equation}
Finally, for the composition of these symmetries, we can simply take the product  $W_{\text{com}}=W_{\text{fpf}}W_{\text{cc}}=-W_{\text{fpf}}$. Naturally, the forms of the matrices $W_{\text{cc}}$, $W_{\text{fpf}}$, and $W_{\text{com}}$ given above are not unique. They should be understood as representatives of the corresponding anyonic symmetries, since different $W$-matrices can induce the same action on the anyon group, differing only by redefinitions involving local particles.

\begin{figure}
	\includegraphics[scale=.65,angle=90]{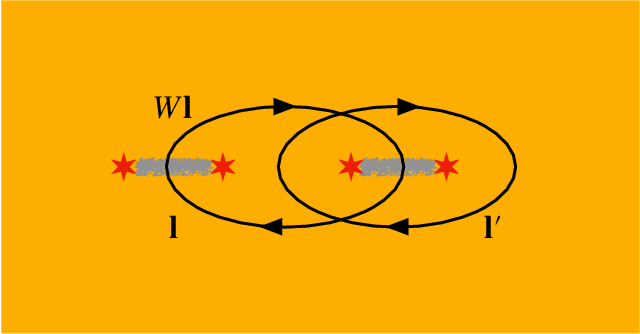}
	\caption{The gray cut lines connect a pair of defects, represented as red triangles. A quasiparticle ${\bf l}$ crossing the cut line gets transmuted into $W{\bf l}$.} 
	\label{DefectsGSD}
\end{figure}

Topological phases endowed with anyonic symmetries support extrinsic defects, which are semi-classical objects that are, in general, non-Abelian \cite{Barkeshli:2013yta,Teo_2016}. When a quasiparticle is transported around such a point-like defect, the quasiparticle gets transmuted according to the underlying anyonic symmetry. The presence of extrinsic defects affects the topological properties of the phase. For example, the presence of two pairs of defects enables a new process that keeps the system in the ground state, as shown in Fig. \ref{DefectsGSD}. This increases the ground state degeneracy compared to the case where the defects are absent. The study of extrinsic defects provides a viable path for the concrete realization of exotic non-Abelian particles.


\section{Interface and Folding \label{IV}}

\begin{figure}
	\includegraphics[scale=.7,angle=90]{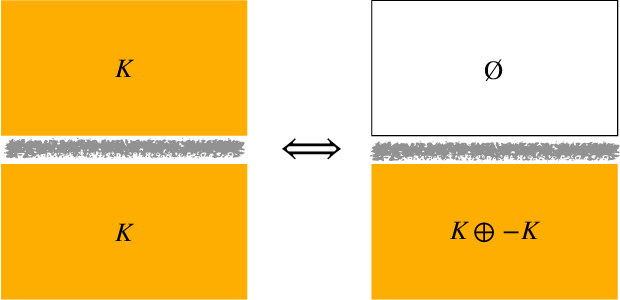}
	\caption{The interface between the topological phases is displayed in gray. In the right, it is shown the folding process.} 
	\label{folding}
\end{figure}

A process like in Fig. \ref{DefectsGSD} can be explicitly realized in a setup involving the interface between two identical topological phases arranged side by side, as depicted in Fig. \ref{folding}. The idea is that by tuning specific interactions at the interface, anyons tunneling through it can be transmuted according to the anyonic symmetries of the topological phase. An alternative way to view such a configuration is by means of the folding procedure \cite{Barkeshli:2013yta}, as illustrated on the right configuration in Fig. \ref{folding}. In this setup, tunneling of particles  is understood in terms of condensation.

Let us start by describing the boundary physics of a single topological phase through the bulk-edge correspondence. For each bulk line operator \eqref{wl}, there is a corresponding vertex operator at the boundary, which can be obtained from semi-infinite lines ending at the boundary  
\begin{equation}
e^{i \int_{-\infty}^0 dy\, l_I \,a_I^y} ~~\Rightarrow ~~ e^{i l_I \phi_I}.
\label{vo}
\end{equation}
The commutation relations of the gauge fields following from the $K$-matrix Chern-Simons theory lead to the boundary commutation relations for the chiral fields
\begin{equation}
[\phi_I(x), \phi_J(x') ]=2\pi i K_{IJ}^{-1} \Theta(x-x').
\label{001}
\end{equation} 

For the case of the system in Fig. \ref{folding}, there are two bulk topological phases sharing the same boundary, through which particles can tunnel. Consequently, there are additional fields both in the bulk and at the boundary. The extra boundary fields satisfy
\begin{equation}
	[\tilde{\phi}_I(x), \tilde{\phi}_J(x') ]=-2\pi i K_{IJ}^{-1} \Theta(x-x').
	\label{002}
\end{equation}
The relations \eqref{001}  and \eqref{002} are encoded into the boundary action
\begin{equation}
	S= \int dx dt \,\frac{1}{4\pi} K_{IJ}\partial_x \phi_I \partial_t\phi_J -  \frac{1}{4\pi} K_{IJ}\partial_x \tilde{\phi}_I \partial_t\tilde{\phi}_J + \cdots,
\end{equation}
where the dots account for interactions, which are dictated by the symmetries. 

Part of the bulk 1-form symmetries becomes 0-form symmetries at the boundary, whose charged operators are the chiral fields
\begin{eqnarray}
\phi_I &\rightarrow& \phi_I +2\pi K^{-1}_{IJ}l_J\nonumber\\
\tilde{\phi}_I &\rightarrow& \tilde{\phi}_I + 2\pi K^{-1}_{IJ}\tilde{l}_J,
\end{eqnarray}
with $l_I, \tilde{l}_I \in \mathbb{Z}^3$. The vertex operators invariant under these symmetries are of the form
\begin{equation}
	e^{ i K_{IJ}\phi_J } ~~~\text{and}~~~  	e^{ i K_{IJ}\tilde{\phi}_J }, 
\end{equation} 
which are either bosons or fermions. Under an anyonic symmetry specified by $W$, these objects transform as 
\begin{eqnarray}
	e^{ i K_{IJ}\phi_J }~~&\overset{W}{\rightarrow}&	~~e^{ i (KW^T)_{IJ}\phi_J }\nonumber\\
		e^{ i K_{IJ}\tilde{\phi}_J }~~&\overset{W}{\rightarrow}&	~~e^{i (KW^T)_{IJ}\tilde{\phi}_J }.
	\label{trans}
\end{eqnarray}

The idea is to consider tunneling events in which particles are transmuted as they cross the interface, namely, events characterized by the nonvanishing amplitude
\begin{equation}
	\big{\langle} e^{i K_{IJ}\phi_J } e^{ i (KW^T)_{IJ}\tilde{\phi}_J }  \big{\rangle}.
\end{equation}
Physically, this means that the interface plays the role of a cut line emanating from symmetry defects. Such tunneling events are enforced by interactions of the form
\begin{equation}
S_{\text{int}} =  \int dx dt \, g_I \cos[K_{IJ}(\phi_J + W^T_{JJ'}\tilde{\phi}_{J'})].
	\label{int+}
\end{equation}
These interactions are invariant under the continuum symmetry
\begin{equation}
\phi_I \rightarrow \phi_I + (W^T \alpha)_I ~~~\text{and}~~~ \tilde{\phi}_I \rightarrow \tilde{\phi}_I - \alpha_I,  
\end{equation}
where $\alpha_I \sim \alpha_I +2\pi$. In addition, note that 
\begin{equation}
	[K_{IJ}(\phi_J + W^T_{JJ'}\tilde{\phi}_{J'}), K_{MN}(\phi_N + W^T_{NN'}\tilde{\phi}_{N'})]=0,~~~\forall~ I, M,
\end{equation}
so that the interactions in  \eqref{int+} satisfy the Haldane criterion \cite{PhysRevLett.74.2090}, and then are able to open a gap in the boundary.

For $W=\openone$, the interactions \eqref{int+} correspond simply to backscattering, where the particles tunnel without transmutation. This type of operator will have an important role in the discussion of the zero modes.

To proceed, it is convenient to introduce the fields
\begin{equation}
	2\theta_I \equiv \phi_I + W^T_{I J} \tilde{\phi}_J~~~\text{and}~~~ 2\varphi_I \equiv \phi_I - W^T_{I J} \tilde{\phi}_J^L,
	\label{redefinition}
\end{equation}
which satisfy
\begin{equation}
	[\theta_I(x),\varphi_J(x')]= i \pi K^{-1}_{IJ}\Theta(x-x').
	\label{basic}
\end{equation}
The inverse relations are
\begin{equation}
	\phi_I = \varphi_I + \theta_I~~~\text{and}~~~ \tilde{\phi}_I= -(W^T)^{-1}_{IJ} (\varphi_J-\theta_J).
\end{equation}
The fields $\theta_I$ and $\varphi_I$ are also $2\pi$-periodic.

In terms of the new fields, the interactions in \eqref{int+} become
\begin{equation}
	S_{\text{int}} =  \int dx dt \, g_I \cos(2 K_{IJ}\theta_J)
	\label{int+1}
\end{equation}
and the full action is
\begin{equation}
	S=\int dx dt\, - \frac{1}{\pi} K_{IJ}\partial_t \theta_I \partial_x\varphi_J + g_I\cos(2 K_{IJ}\theta_J)+\cdots,
	\label{full}
\end{equation}
where the dots incorporate forward interactions. 

\begin{figure}
	\includegraphics[scale=.6,angle=90]{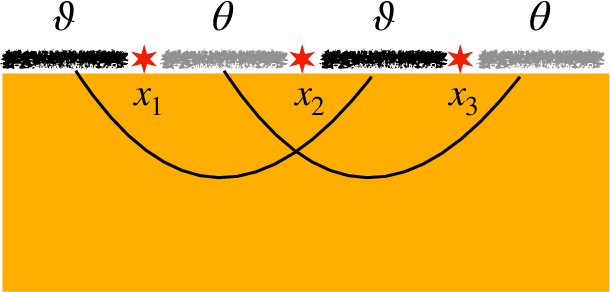}
	\caption{Configuration with $2N=4$ alternating (black and gray) regions supporting zero modes, localized at the positions marked with red triangles. The gray part of the boundary corresponds to the region where anyons differing by the action of the symmetry $W$ (transmutation) are condensed, whereas the black region the condensed anyons are identical (backscattering). The lines represent the process of Fig. \ref{DefectsGSD}.} 
	\label{zeromodes}
\end{figure}

The boundary theory \eqref{full} enforces the transmutation of particles as they cross the interface. However, to localize zero modes, we need a slightly more intricate setup that mimics the configuration of Fig. \ref{DefectsGSD} \cite{Barkeshli:2012pr,Barkeshli_2012,Barkeshli:2013yta}. It also involves gapping terms which do not realize any transmutation as the particles cross the interface, alternating with the ones in \eqref{full}. The setup, from the folding perspective, is shown in Fig. \ref{zeromodes} and the corresponding action is 
\begin{equation}
	S=\int dx dt\, - \frac{1}{\pi} K_{IJ}\partial_t \theta_I \partial_x\varphi_J + g_I(x)\cos(2 K_{IJ}\theta_J)+h_I(x)\cos(2 K_{IJ}\vartheta_J)+\cdots,
	\label{full1}
\end{equation}
where we have defined 
\begin{equation}
	2\vartheta_I \equiv 2\theta_I\big{|}_{W=\openone} =\phi_I + \tilde{\phi}_I.
	\label{redefinition2}
\end{equation}
The coupling constants $g_I(x)$ are nonvanishing only in the gray regions of Fig. \ref{zeromodes}, whereas $h_I(x)$ are nonvaishing only in the black regions.

To proceed, we compute the algebra between operators:
\begin{equation}
e^{i \theta_I(x)} e^{i \vartheta_J(x')}= \exp[-\frac{\pi i}{2}\Theta(x-x') [(\openone-W^{T})K^{-1}]_{IJ}]\  e^{i \vartheta_J(x')}e^{i \theta_I(x)}
\label{fullalgebra}
\end{equation}
and
\begin{equation}
	e^{i \vartheta_I(x)} e^{i \varphi_J(x')}= \exp[-\frac{\pi i}{2}\Theta(x-x') [(\openone+W^{T})K^{-1}]_{IJ}]\  e^{i \varphi_J(x')}e^{i \vartheta_I(x)}.
	\label{fullalgebra5}
\end{equation}
These relations are important in analyzing the ground state degeneracy of the boundary theory, particularly in understanding which contributions arise from the anyonic bulk symmetry. 

To disentangle bulk and boundary contributions, we consider the operators which are associated with the bulk Wilson lines, 
\begin{equation}
	e^{i\int_{-\infty}^0 (l_I a_I+ \tilde{l}_I \tilde{a}_I )} ~~\Rightarrow ~~ e^{i (l_I \phi_I+\tilde{l}_I \tilde\phi_I)}.
\end{equation}
The simplest operators expressed in terms of $\theta$, $\varphi$, and $\vartheta$ correspond to setting $l_I=\delta_{JI}$ and $\tilde{l}_I =\pm W^T_{JI}$, 
\begin{equation}
	e^{i ( \phi_I+W^T_{IJ} \tilde\phi_J)}=e^{2 i \theta_I}~~~\text{and}~~~	e^{i ( \phi_I-W^T_{IJ} \tilde\phi_J)}=e^{2 i \varphi_I},
	\label{556}
\end{equation}
and setting $l_I=\tilde{l}_I=\delta_{JI}$, 
\begin{equation}
	e^{i ( \phi_I+\tilde\phi_I)}=e^{2 i \vartheta_I}.
	\label{557}
\end{equation}
We see that all operators associated with the bulk Wilson lines have even factors in the exponential. We refer to $e^{2 i \theta_I}$, $e^{2 i \varphi_I}$, and $e^{2 i \vartheta_I}$ as Wilson operators. In particular, the boundary counterpart of the nontrivial commutation between the Wilson lines in Fig. \ref{zeromodes} reads
\begin{equation}
	e^{2i \theta_I(x)} e^{2i \vartheta_J(x')}= \exp[-2\pi i\Theta(x-x') [(\openone-W^{T})K^{-1}]_{IJ}]\  e^{2i \vartheta_J(x')}e^{2i \theta_I(x)},
	\label{fullalgebra45}
\end{equation}
which accounts for the contribution to the ground state degeneracy originating from the defects associated with the anyonic symmetries. This degeneracy is robust. 
It persists even when the system interacts with an external environment that breaks all the intrinsic boundary symmetries, provided that the bulk remains topologically ordered.

We emphasize that twist defects are, in general, defined intrinsically in the bulk through their action on anyon labels, namely through the corresponding anyonic symmetry~\cite{Barkeshli:2012pr}. In the present formulation, this bulk relabeling is captured by Eq.~\eqref{fullalgebra45}. The additional zero modes that appear in the boundary realization should instead be understood as contributions associated with domain walls localized at the boundary. Consequently, the quantum dimensions extracted from the algebra of Wilson operators are independent of the particular $K$-matrix representative chosen for a given topological order, as well as of the particular $W$-matrix representative chosen for a given anyonic symmetry. They are determined only by the bulk anyonic data and the induced action of the anyonic symmetry on the anyon group.

The folded picture generally admits gapped boundaries, classified by Lagrangian subgroups, that do not arise from bulk anyonic symmetries. Their junctions may support additional zero modes that are not associated with permutations of the bulk anyons. In this work, however, we focus specifically on the subset of gapped boundaries associated with invertible bulk anyonic symmetries, for which the correspondence between the folded and unfolded descriptions applies directly. We provide an explicit example illustrating this distinction in Sec.~\ref{conclusions}.

\section{Non-Abelian Zero Modes \label{V}}

We discuss now the explicit construction of the non-Abelian zero modes associated with the anyonic symmetries. The nature of the zero modes depends on which symmetry is favored by the gapping terms in \eqref{full1} (through the fields $\theta_I$). One point we need to be careful about is that the total ground state degeneracy receives contributions not only from the anyonic symmetries but also from the intrinsic boundary physics. We shall discuss how these contributions can be identified and split.

In the IR, the Hamiltonian is dominated by 
\begin{eqnarray}
H_{IR} \sim &-& \int dx \left[g_1(x) \cos(24 \theta_1)+g_2(x) \cos(2 \theta_2)+g_3(x) \cos(2 \theta_3)\right] \nonumber\\
&-& \int dx \left[h_1(x) \cos(24 \vartheta_1)+h_2(x) \cos(2 \vartheta_2)+h_3(x) \cos(2 \vartheta_3)\right],
\end{eqnarray}
which pins the fields at the minima of the cosines. We proceed to find the ground state degeneracy. As it can be checked explicitly from the algebras presented below, all the operators $e^{i\theta}$'s, $e^{i\varphi}$'s, and $e^{i\vartheta}$'s do commute with the Hamiltonian but do not commute in general with each other. Therefore, their algebras form a  ground state representation up to exponentially small corrections\footnote{We are assuming that the length of each one of the alternating regions in Fig. \ref{zeromodes} is large enough so that the energy splitting between the states is negligible.}. The specific commutation relations depend on the $W$-matrix and we shall discuss each case separately. 

At this point, we choose to diagonalize the simultaneously commuting operators
\begin{eqnarray}
e^{i \theta_1} \ket{n_1} &=& e^{2\pi i\frac{n_1}{24} }\ket{n_1},~~~ n_1=0,1,\ldots,23, \nonumber\\
e^{i \theta_2} \ket{n_2} &=& e^{2\pi i \frac{n_2}{2} }\ket{n_2},~~~ n_2=0,1, \nonumber\\
e^{i \theta_3} \ket{n_3} &=& e^{2 \pi i \frac{n_3}{2} }\ket{n_3},~~~ n_3=0,1.
\label{diagonal}
\end{eqnarray}
We need to find which operators can be diagonalized along with $e^{i\theta}$'s. We first note that, according to the commutations in \eqref{basic}, the operators $e^{i\varphi}$'s act as ladder operators for the above eigenstates with unit step size, spanning the whole spectrum of $e^{i\theta}$'s. To further proceed, we have to study the commutations involving $e^{i\vartheta}$'s.


\subsubsection{Charge Conjugation}

Let us consider initially that the boundary gapping terms in \eqref{full1} involve the charge conjugation $W$-matrix given in \eqref{cc}.  Note that in this case, $\varphi_I=\vartheta_I$. According to \eqref{fullalgebra}, it follows that
\begin{eqnarray}
e^{i \theta_1(x)}  e^{i \vartheta_1(x')} &=& e^{-\frac{\pi i}{12}\Theta(x-x') } e^{i \vartheta_1(x')} e^{i \theta_1(x)},\nonumber\\
e^{i \theta_2(x)}  e^{i \vartheta_2(x')} &=& e^{-\pi i\Theta(x-x') }  e^{i \vartheta_2(x')} e^{i \theta_2(x)},\nonumber\\
e^{i \theta_3(x)}  e^{i \vartheta_3(x')} &=& e^{\pi i\Theta(x-x') }  e^{i \vartheta_3(x')} e^{i \theta_3(x)},
\label{oacc}
\end{eqnarray}
with all other commutation relations being trivial. The operators which can be diagonalized along with the ones in \eqref{diagonal} are $e^{24 i \vartheta_1},  e^{2 i \vartheta_2}$, and $e^{2 i \vartheta_3}$. Each of these operators has a single associated eigenstate, so that the total number of boundary states follows exclusively from $\ket{n_1,n_2,n_3}$ in \eqref{diagonal}, providing  $24 \times 2\times 2$ states. All these states are connected through the algebra  in \eqref{oacc}, since $e^{i\vartheta}=e^{i\varphi}$ act as ladder operators with unit step size. To unveil the quantum dimension of the associated defects, we need to find the scaling of the ground state degeneracy as we 
consider a system with $2N$ alternating regions as in Fig. \ref{zeromodes}. This leads to $24^N \times 2^N\times 2^N$ states, from which we find that the corresponding defect has quantum dimension $\sqrt{24 \times 2\times 2}$.

Part of this degeneracy comes from the bulk physics, and it is dictated by the quantum dimension of the defect associated with charge conjugation while the remaining contributions are purely due to the boundary physics. To identify these contributions, we study the algebra of the Wilson operators \eqref{556} and \eqref{557}. The only nontrivial commutation relation is 
\begin{equation}
e^{2 i \theta_1(x)}  e^{2 i \vartheta_1(x')} = e^{-\frac{\pi i}{3}\Theta(x-x') } e^{2 i \vartheta_1(x')} e^{2 i \theta_1(x)}.
\label{WilsonAlgebraCC}
\end{equation}
The operator $e^{2 i \theta_1}$ has 12 eigenstates but not all of them are accessible since the above algebra implies that the ladder operator $e^{2 i \vartheta_1} $
has step size 2, so that we end up with $12/2 =6$ accessible states. Thus, from $4 \times 6$ eigenstates of $e^{i \theta_1}$, we see that the factor of 4 is purely from the edge physics whereas the factor 6 is a consequence of the anyonic symmetry. Thus, the quantum dimension of the corresponding defect is $\sqrt{6}$.  
Also, $e^{2 i \theta_2}$ and $e^{2 i \theta_3}$ have unique eigenstates, which means that the degeneracy coming from $e^{ i \theta_2}$ and $e^{ i \theta_3}$ is purely from edge physics. This intrinsic boundary degeneracy corresponds to the usual degeneracy in one-dimensional fermionic systems supporting Majorana zero modes of dimension $\sqrt{2}$ \cite{Kitaev_2001,PhysRevB.83.075103,PhysRevB.84.195436}.

By following \cite{Barkeshli:2013yta}, we can construct explicitly the zero modes associated with the anyonic symmetry as
\begin{eqnarray}
\gamma_i^{\text{cc}}\equiv \lim_{\epsilon \rightarrow 0^+ }
	\begin{cases}
		e^{2 i \vartheta_1(x_i -\epsilon)}e^{2 i \theta_1(x_i +\epsilon)},\quad \text{if } i~ \text{is odd},\\
		e^{2 i \theta_1(x_i -\epsilon)} e^{2 i \vartheta_1(x_i +\epsilon)}, \quad \text{if } i~ \text{is even},
	\end{cases}
\label{cczeromode}
\end{eqnarray}
which are expressed in terms of Wilson operators (operators with even factors in the exponentials). They represent zero modes exponentially localized in the domain walls at the positions $x_i$'s, as shown in Fig. \ref{zeromodes} (red triangles). From \eqref{WilsonAlgebraCC}, we obtain that they satisfy 
\begin{equation}
\gamma_i^{\text{cc}}\gamma_j^{\text{cc}}= e^{-\frac{\pi i }{3} \text{sgn}(x_i-x_j)}\gamma_j^{\text{cc}} \gamma_i^{\text{cc}}, 
\label{cc00}
\end{equation}
which shows that $\gamma_i^{\text{cc}}$'s are $\mathbb{Z}_6$ parafermions, as expected by defects of quantum dimension $\sqrt{6}$.

Now we couple the boundary to an external system in such a way that the only remaining symmetries are the shifts following from the algebra of Wilson operators  \eqref{WilsonAlgebraCC}. In this situation, only the zero modes \eqref{cc00} will manifest. The low-energy effective Hamiltonian retaining only the symmetries associated with \eqref{WilsonAlgebraCC} is
\begin{equation}
	H_{\text{cc}} \sim -\int dx \left[ \tilde{g}_1(x) \cos[6(2\theta_1)]+\tilde{h}_1(x) \cos[6(2\vartheta_1)] \right],
	\label{effcc}
\end{equation}
where $ \tilde{g}_1(x)$ is nonvanishing only in the gray regions of Fig. \ref{zeromodes}, whereas $\tilde{h}_1(x)$ is nonvanishing only in the black regions. We can check that $\left[H_{\text{cc}} , \gamma_i^{\text{cc}}\right]=0$, as it should be. The effective Hamiltonian \eqref{effcc} will be used later to explore the physical signatures of the zero modes.


\subsubsection{Fermion Parity Flip}

We now examine the case of interactions favoring the fermion parity flip symmetry, whose $W$-matrix in \eqref{parity} leads to the following nontrivial commutation relations
\begin{eqnarray}
	e^{i \theta_1(x)}  e^{i \vartheta_1(x')} &=& e^{\frac{\pi i}{4}\Theta(x-x') } e^{i \vartheta_1(x')} e^{i \theta_1(x)},\nonumber\\
		e^{i \theta_1(x)}  e^{i \vartheta_3(x')} &=& e^{-\pi i\Theta(x-x') } e^{i \vartheta_3(x')} e^{i \theta_1(x)},\nonumber\\
	e^{i \theta_2(x)}  e^{i \vartheta_2(x')} &=& e^{-\pi i\Theta(x-x') }  e^{i \vartheta_2(x')} e^{i \theta_2(x)},\nonumber\\
	e^{i \theta_3(x)}  e^{i \vartheta_1(x')} &=& e^{-\pi i\Theta(x-x') }  e^{i \vartheta_1(x')} e^{i \theta_3(x)}.
	\label{oafpf}
\end{eqnarray}
The operators that can be simultaneously diagonalized along with the ones in \eqref{diagonal} are $e^{8i \vartheta_1}$, $e^{2i \vartheta_2}$, and $e^{2i \vartheta_3}$. While the operators $e^{2i \vartheta_2}$ and $e^{2i \vartheta_3}$ have single eigenstates, the operator $e^{8i \vartheta_1}$ has three eigenstates, which we specify as 
\begin{equation}
e^{8i \vartheta_1}\ket{m_1} =  e^{2\pi i\frac{m_1}{3} }\ket{m_1},~~~ m_1=0,1,2.
\end{equation}
Thus, the complete basis of states in the case of fermion parity flip is $\ket{n_1,n_2,n_3,m_1}$. 

As in the previous case, we can also use here the operators $e^{i\varphi}$'s to run over the states. In particular, the relation  
\begin{equation}
e^{i \vartheta_1(x)}  e^{i \varphi_1(x')} = e^{-\frac{\pi i}{3}\Theta(x-x') } e^{i \varphi_1(x')} e^{i \vartheta_1(x)},
\end{equation}
implies that $e^{i\varphi_1}$ is also a ladder operator for the states $\ket{m_1}$ with step size 1, in addition to being a ladder operator for $\ket{n_1}$. In this way,  the eigenvalues $n_1$ and $m_1$ are not independent, resulting in $24$ states. The operator $e^{i \varphi_2} $ is a ladder operator for the eigenstates of $e^{i \theta_2} $, with step size 1; likewise $e^{i \varphi_3} $ is a ladder operator for the eigenstates of $e^{i \theta_3} $. Therefore, the total number of states is  $24\times 2\times 2$. When we consider $2N$ alternating regions, the ground state degeneracy scales as $24^N\times2^N \times 2^N$.

To find the contribution coming from the fermion parity flip symmetry to this degeneracy, we consider the Wilson operators $e^{2 i \theta_1(x)}$ and $e^{4 i \vartheta_1(x)}$, which can be simultaneously diagonalized, 
\begin{eqnarray}
e^{2 i \theta_1(x)} \ket{n_1'}&=&e^{2\pi i \frac{n_1'}{12}}  \ket{n_1'},~~~n_1'=0,1,\ldots,11,\nonumber\\
e^{4 i \vartheta_1(x)} \ket{m_1'}&=&e^{2\pi i \frac{m_1'}{6}}  \ket{m_1'},~~~m_1'=0,1,\ldots,5.
\end{eqnarray}
We can use $e^{2i\varphi_1}$ as the ladder operator for such states, 
\begin{eqnarray}
e^{2i\theta_1(x)} e^{2i\varphi_1(x')} &=& e^{-\frac{\pi i}{3}\Theta(x-x')}  e^{2i\varphi_1(x')} e^{2i\theta_1(x)},\nonumber\\
	e^{4i\vartheta_1(x)} e^{2i\varphi_1(x')} &=& e^{-\frac{8 \pi i}{3}\Theta(x-x')} e^{2i\varphi_1(x')} e^{4i\vartheta_1(x)}. 
\end{eqnarray}
These relations imply that $e^{2i\varphi_1}$ is a simultaneous ladder operator for both states $\ket{n_1'}$ and $\ket{m_1'}$ with step size two for both. Therefore, the total number of states specified by $\ket{n_1', m_1'}$ is 6. It seems that the number of states associated with the fermion parity flip is the same as in the case of charge conjugation, but that is not quite correct. In fact, the algebra of the zero modes is dictated by the relation, 
\begin{equation}
e^{2 i \theta_1(x)}  e^{2 i \vartheta_1(x')} = e^{\pi i\Theta(x-x') } e^{2 i \vartheta_1(x')} e^{2 i \theta_1(x)},
\label{9908}
\end{equation}
which implies that the operator $e^{2 i \vartheta_1}$ is a ladder operator for the states $\ket{n_1'}$ with step size 6, without affecting the states $\ket{m_1'}$. The corresponding zero modes are
\begin{eqnarray}
	\gamma_i^{\text{fpf}}\equiv \lim_{\epsilon \rightarrow 0^+ }
	\begin{cases}
		e^{2 i \vartheta_1(x_i -\epsilon)}e^{2 i \theta_1(x_i +\epsilon)},\quad \text{if } i~ \text{is odd},\\
		e^{2 i \theta_1(x_i -\epsilon)} e^{2 i \vartheta_1(x_i +\epsilon)}, \quad \text{if } i~ \text{is even}.
	\end{cases}
	\label{fpfzeromode}
\end{eqnarray}
They satisfy 
\begin{equation}
	\gamma_i^{\text{fpf}}\gamma_j^{\text{fpf}}= e^{\pi i \, \text{sgn}(x_i-x_j)}\gamma_j^{\text{fpf}} \gamma_i^{\text{fpf}},
	\label{fpf00}
\end{equation}
showing that they are Majorana modes. So, from the 6 states specified by $ \ket{n_1',m_1'}$, we see that only 2 of them come from the fermion parity flip symmetry, and these two states are connected through the application of $e^{2 i \vartheta_1}$.

As we did before, we couple the boundary to an external system which breaks all the intrinsic boundary symmetries, leaving only the shift symmetries resulting from \eqref{9908}. The low-energy effective Hamiltonian incorporating this feature is
\begin{equation}
	H_{\text{fpf}} \sim -\int dx \left[ \tilde{g}_1(x) \cos[2(2\theta_1)]+\tilde{h}_1(x) \cos[2(2\vartheta_1)] \right],
	\label{fpfef}
\end{equation}
where $ \tilde{g}_1(x)$ and $\tilde{h}_1(x)$ are defined as before. This Hamiltonian hosts exclusively the zero modes in \eqref{fpfzeromode}, $[H_{\text{fpf}}, \gamma_i^{\text{fpf}}]=0$.


\subsubsection{Composite Symmetry}

Finally, in the case where the gapping terms favor the composite symmetry, whose $W$-matrix is $W_{\text{com}}=W_{\text{fpf}}W_{\text{cc}}=-W_{\text{fpf}}$, we have the nontrivial relations
\begin{eqnarray}
	e^{i \theta_1(x)}  e^{i \vartheta_1(x')} &=& e^{-\frac{\pi i}{3}\Theta(x-x') } e^{i \vartheta_1(x')} e^{i \theta_1(x)},\nonumber\\
	e^{i \theta_1(x)}  e^{i \vartheta_3(x')} &=& e^{\pi i\Theta(x-x') } e^{i \vartheta_3(x')} e^{i \theta_1(x)},\nonumber\\
	e^{i \theta_3(x)}  e^{i \vartheta_1(x')} &=& e^{\pi i\Theta(x-x') }  e^{i \vartheta_1(x')} e^{i \theta_3(x)},\nonumber\\
	e^{i \theta_3(x)}  e^{i \vartheta_3(x')} &=& e^{-3 \pi i\Theta(x-x') }  e^{i \vartheta_3(x')} e^{i \theta_3(x)}.
	\label{oacs}
\end{eqnarray}
The operators that commute with the ones in  \eqref{diagonal} are $e^{6i\vartheta_1}$, $e^{i\vartheta_2}$, and $e^{2i\vartheta_3}$. While this last one has a single eigenstate, the remaining operators have more states, namely, 
\begin{eqnarray}
e^{6i \vartheta_1}\ket{m_1} &=&  e^{2\pi i\frac{m_1}{4} }\ket{m_1},~~~ m_1=0,\ldots,3,\nonumber\\
e^{i \vartheta_2}\ket{m_2} &=&  e^{2\pi i\frac{m_2}{2} }\ket{m_2},~~~ m_2=0,1.
\end{eqnarray}
The complete basis is therefore $\ket{n_1,n_2,n_3,m_1,m_2}$. The operators $e^{i\varphi}$'s are ladder operators for such states. In particular, as $e^{i\varphi_1}$ acts nontrivially on the eigenstates $\ket{n_1}$ and  $\ket{m_1}$, the corresponding eigenvalues are not independent. The same goes for the eigenstates $\ket{n_2}$ and  $\ket{m_2}$, since $e^{i\varphi_2}$ acts nontrivially on both. The total number of states is again $24\times 2 \times 2$. The ground state degeneracy scales as $24^N\times 2^N \times 2^N$ for a system with $2N$ alternating regions.

To find the contribution from the composite symmetry to this degeneracy, we consider the algebra of Wilson operators $e^{2 i \theta_1(x)}$ and $e^{6 i \vartheta_1(x)}$, which can be simultaneously diagonalized, 
\begin{eqnarray}
	e^{2 i \theta_1(x)} \ket{n_1'}&=&e^{2\pi i \frac{n_1'}{12}}  \ket{n_1'},~~~n_1'=0,1,\ldots,11,\nonumber\\
	e^{6 i \vartheta_1(x)} \ket{m_1'}&=&e^{2\pi i \frac{m_1'}{4}}  \ket{m_1'},~~~n_1'=0,1,\ldots,3.
\end{eqnarray}
We can use $e^{2i\varphi_1}$ as the ladder operator for such states, 
\begin{eqnarray}
	e^{2i\theta_1(x)} e^{2i\varphi_1(x')} &=& e^{-\frac{\pi i}{3}\Theta(x-x')}  e^{2i\varphi_1(x')} e^{2i\theta_1(x)},\nonumber\\
	e^{6i\vartheta_1(x)} e^{2i\varphi_1(x')} &=& e^{3 \pi i\Theta(x-x')} e^{2i\varphi_1(x')} e^{4i\vartheta_1(x)}. 
\end{eqnarray}
These relations imply that $e^{2i\varphi_1}$ is a simultaneous ladder operator for both states $\ket{n_1'}$ and $\ket{m_1'}$ with step size two for both. Therefore, the total number of states specified by $\ket{n_1', m_1'}$ is 6. 

As in the case of the fermion parity flip, not all these states are associated with the anyonic symmetry. To find the states corresponding to the composite symmetry, we consider
\begin{equation}
e^{2i \theta_1(x)}  e^{2i \vartheta_1(x')} = e^{-\frac{4\pi i}{3}\Theta(x-x') } e^{2i \vartheta_1(x')} e^{2i \theta_1(x)},
\label{9987}
\end{equation}
which implies that $e^{2i \vartheta_1}$ is a ladder operator for the twelve eigenstates of $e^{2i \theta_1}$ with step size 4, so that we get $3$ states.  The algebra \eqref{9987} implies that the zero modes 
\begin{eqnarray}
	\gamma_i^{\text{com}}\equiv \lim_{\epsilon \rightarrow 0^+ }
	\begin{cases}
		e^{2 i \vartheta_1(x_i -\epsilon)}e^{2 i \theta_1(x_i +\epsilon)},\quad \text{if } i~ \text{is odd},\\
		e^{2 i \theta_1(x_i -\epsilon)} e^{2 i \vartheta_1(x_i +\epsilon)}, \quad \text{if } i~ \text{is even},
	\end{cases}
	\label{comzeromode}
\end{eqnarray}
satisfy
\begin{equation}
	\gamma_i^{\text{com}}\gamma_j^{\text{com}}= e^{-\frac{4\pi i }{3} \text{sgn}(x_i-x_j)}\gamma_j^{\text{com}} \gamma_i^{\text{com}},
	\label{com00}
\end{equation}
showing that there are $\mathbb{Z}_3$ parafermions. It is interesting to note that these defects are closely related to those associated with charge conjugation anyonic symmetry in level-$k$ FQH systems. Further details on the relation between the composite-symmetry defects in the level-$4k$ theory and the charge conjugation defects in the level-$k$ theory are provided in Appendix~\ref{appe}.

Breaking the intrinsic boundary symmetries through the coupling to an external system,  we can write the low-energy effective Hamiltonian as 
\begin{equation}
	H_{\text{com}} \sim -\int dx \left[ \tilde{g}_1(x) \cos[3(2\theta_1)]+\tilde{h}_1(x) \cos[3(2\vartheta_1)] \right],
	\label{comeff}
\end{equation}
which is invariant under the shifts implied by \eqref{9987} and, consequently, supports the zero modes \eqref{comzeromode}, $\left[H_{\text{com}} , \gamma_i^{\text{com}}\right]=0$.


\section{Signatures of Zero Modes via Josephson Effect \label{VI}}

The presence of non-Abelian zero modes manifests in the periodicity of Josephson currents \cite{Kitaev_2001,Kwon_2003,Fu_2009,Clarke_2013,Cheng:2012xh,Fu_2009,Beenakker_2013,Kwon_2003,Alicea2011bjx}. To explore this point, let us consider a particular region in Fig. \ref{zeromodes}, say the region $x_3-x_2$, whose gapping terms do not favor any anyonic symmetry (they are of the form $\cos(2 K_{IJ}\vartheta_J)$). We then take the Josephson effect regime, where the length $x_3-x_2$ is much smaller than the lengths of the adjacent regions, i.e.,  $x_3-x_2 \ll x_2-x_1 \sim x_4-x_3$, in order to allow tunneling current.

Next, we consider the low-energy effective Hamiltonians \eqref{effcc}, \eqref{fpfef}, and \eqref{comeff} in the Josephson effect regime. We can focus on
\begin{equation}
H_{\text{Josephson}}^a= - \int_{x_1}^{x_2}dx\, \tilde{g}_1(x) \cos [\mathsf{d}_a (2\theta_1)] - \int_{x_3}^{x_4}dx\, \tilde{g}_1(x) \cos [\mathsf{d}_a (2\theta_1)+\delta\theta_a], 
\label{Josepshon}
\end{equation}
where we describe the three cases in a unified way, i.e., the index $a$ corresponds to $a= \text{cc}, \text{fpf}, \text{com}$. The factor $\mathsf{d}_a$ is the degeneracy associated with the anyonic symmetries, $\mathsf{d}_{\text{cc}} = 6$, $\mathsf{d}_{\text{fpf}} = 2$, and $\mathsf{d}_{\text{com}}= 3$, and $\delta\theta_a$ are the corresponding phase differences.

The tunneling current is in general dominated by the zero modes. Thus, the low-energy effective tunneling Hamiltonians can be expressed as
\begin{equation}
H_{\text{tunn}}^a= - t \left( \gamma_2^{a \dagger} \gamma_3^a +\,\text{H. c.} \right),
\label{t}
\end{equation}
where the zero modes are given by
\begin{equation}
\gamma_2^{a}=  \lim_{\epsilon \rightarrow 0^+ } e^{2 i \theta_1(x_2-\epsilon)} e^{2 i \vartheta_1(x_2+\epsilon)}
\end{equation}
and
\begin{equation}
\gamma_3^{a}=  \lim_{\epsilon \rightarrow 0^+ } e^{2 i \vartheta_1(x_3-\epsilon)} e^{2 i \theta_1(x_3+\epsilon) + i\frac{\delta\theta_a}{\mathsf{d}_a}},
\end{equation}
with $\gamma_3^{a}$ taking into account the phase difference dictated by \eqref{Josepshon}. Therefore, we see that the phase periodicity of the tunneling Hamiltonian \eqref{t} is 
\begin{equation}
\delta\theta_a \rightarrow \delta\theta_a +2\pi \mathsf{d}_a.
\end{equation}
Writing explicitly for all the cases, 
\begin{eqnarray}
\delta\theta_{\text{cc}} &\rightarrow& \delta\theta_{\text{cc}}  +12\pi,\nonumber\\
\delta\theta_{\text{fpf}} &\rightarrow& \delta\theta_{\text{fpf}}  +4\pi,\nonumber\\	
\delta\theta_{\text{com}} &\rightarrow& \delta\theta_{\text{com}}  +6\pi.	
\end{eqnarray}	 
In particular, in the case of fermion parity flip, whose zero modes are Majoranas, we recover the $4\pi$-periodicity \cite{Kitaev_2001,Kwon_2003,Fu_2009,Fu_2009,Beenakker_2013,Kwon_2003}. In Table~\ref{tab:zero_modes}, we summarize the main properties of the twist defects realizable in the platform considered in this work.

\setlength{\tabcolsep}{4pt} 

\begin{table}[h]
\centering
\begin{tabular}{cccc}
\toprule
\textbf{Symmetry} & \textbf{Quantum Dimension} & \textbf{Zero-Mode Type} & \textbf{Josephson Periodicity} \\
\midrule
Charge Conjugation & $\sqrt{6}$ & $\mathbb{Z}_6$ Parafermion & $12\pi$ \\
Fermion Parity Flip & $\sqrt{2}$ & $\mathbb{Z}_2$ Parafermion & $4\pi$ \\
Composite & $\sqrt{3}$ & $\mathbb{Z}_3$ Parafermion & $6\pi$ \\
\bottomrule
\end{tabular}
\caption{Comparison of topological zero modes.}
\label{tab:zero_modes}
\end{table}


\section{Conclusions}\label{conclusions}

We have addressed the problem of finding concrete realizations of non-Abelian zero modes in Abelian topological  phases. Abelian phases with anyonic symmetries support such non-Abelian modes in the form of the symmetry defects, which can be realized at their boundaries. This can be achieved by considering  two identical phases disposed side by side and adding interactions at the interface favoring the implementation of a specific  anyonic symmetry. 

The phases we study arise from a heterostructure composed of an FQH and a conventional superconductor. The superconductor effectively induces an extra fractionalization of the FQH excitations, so that the resulting spectrum of anyons exhibits new anyonic symmetries. Consequently, such a system  supports different types of zero modes at fixed filling fraction. This naturally raises the question of what type of mode is actually realized at the edge. 

The nature of the zero modes that appear depends on the specific interactions at the boundary. One issue in modeling such interactions is that we end up introducing  intrinsic boundary contributions to the ground state degeneracy,  mingling with the contributions coming from the anyonic symmetries. The strategy is to couple the boundary to an external system which breaks the intrinsic boundary symmetries, remaining only the robust part associated with the symmetry-defects. Then we construct effective Hamiltonians encoding such mechanism and derive the periodicities of the Josephson current, which carry the signatures of the zero modes.

Although we focused on the case $k=3$, the discussion can be generalized without conceptual difficulty for generic $k$, as the nontrivial anyonic symmetries are encoded in the bosonic sector $4k$. Putting aside the question of finding the representative $W$-matrices realizing the symmetries for a given $k$, the quantum dimension of the defects associated  with the discussed symmetries are $\sqrt{\mathsf{d}_a}$, with $\mathsf{d}_{\text{cc}}=2 k$, $\mathsf{d}_{\text{fpf}}=2$, and $\mathsf{d}_{\text{com}}=k$. These defects dictate the periodicity $2\pi \mathsf{d}_a$ of the Josephson current. 

Finally, as already noted at the end of Sec.~\ref{IV}, this generalization does not exhaust all possible gapped boundaries of the folded theory. For example, when $k=9$, the theory with $K=\operatorname{diag}(36,-36)$ has six Lagrangian subgroups. Four correspond to the identity and the three nontrivial invertible anyonic symmetries of $U(1)_{36}$, with defect quantum dimensions $\sqrt{\mathsf{d}_{\text{trivial}}}=1$, $\sqrt{\mathsf{d}_{\text{cc}}}=3\sqrt{2}$, $\sqrt{\mathsf{d}_{\text{fpf}}}=\sqrt{2}$, and $\sqrt{\mathsf{d}_{\text{com}}}=3$. Upon unfolding, the remaining two Lagrangian subgroups describe anyon-condensing interfaces that do not implement permutations of the bulk anyons. Their junctions with the diagonal interface support zero modes with quantum dimensions $\sqrt{3}$ and $\sqrt{6}$. Exploring the boundary realization of these additional zero modes and their possible signatures in the Josephson current is a natural direction for future work.


\begin{acknowledgments}

We would like to thank Carlos Hernaski for his participation in the initial stage of this work. We also thank Claudio Chamon and Rodrigo Pereira for useful discussions. We acknowledge the financial support from the Brazilian funding agencies Fundação Araucária, CAPES, and CNPq. 
	
\end{acknowledgments}


\appendix


\section{Comparison of Defects in FQH+SC and FQH}\label{appe}

The defects associated with the composite symmetry in the level-$4k$ FQH+SC system were found to have quantum dimension $\sqrt{k}$. These defects can be naturally related to the defects associated with charge conjugation anyonic symmetry in the level-$k$ FQH system. This relation becomes transparent by using the decomposition, valid for odd $k$,
\begin{equation}
\mathbb Z_{4k}\cong \mathbb Z_4\times \mathbb Z_k .
\end{equation}
The generators of the $\mathbb Z_4$ and $\mathbb Z_k$ anyon sectors are
\begin{equation}
e^{i k \oint a_1}
\qquad \text{and} \qquad
e^{i 4 \oint a_1},
\end{equation}
respectively. The composite symmetry acts on the respective anyon label as
\begin{equation}
k \overset{\text{com}}{\rightarrow} -k +2k =k ~~~\text{and}~~~ 4 \overset{\text{com}}{\rightarrow} -4.     
\end{equation}
Thus, under the decomposition $\mathbb Z_{4k}\cong \mathbb Z_4\times \mathbb Z_k$, the composite symmetry acts trivially on the $\mathbb Z_4$ sector and as charge conjugation on the $\mathbb Z_k$ sector. This explains why the corresponding defects reproduce the structure of charge conjugation defects in the level-$k$ FQH theory.

\bibliographystyle{ieeetr}
\bibliography{references}

\end{document}